# Effect of Cu(II), Pb(II), Mg(II) ions on γ-irradiated *Spirulina platensis*

E. Gelagutashvili , N.Bagdavadze, D. Jishiashvili, A.Gongadze, M.Gogebashvili, N.Ivanishvili

Iv. Javakhishvili Tbilisi State University, E. Andronikashvili Institute of Physics

0177, 6, Tamarashvili St.,

Tbilisi, Georgia

E.mail:eterige@gmail.com

## Abstract

Influence of metal ions Cu(II), Pb(II), Mg(II) on *Spirulina platensis* cells were studied after three times $Cs^{137}$ gamma irradiation discrete dose (in each case 20 kGy irradiation dose) and recultivation using UV-VIS spectrometry.

It was shown, that metal ions Mg(II), Pb(II), Cu(II) differ from each other in terms of interaction efficiency. In particular, Mg(II) ions act as a stimulant to increase content for *Spirulina platensis* components after several times with discrete doses of radiation and cultivation . For Pb (II) ions, the change in absorption intensity at low concentrations slowly decreases, and then with an increase in the Pb (II) concentration, the absorption intensity increases for C-phycocyanin, Ch1a, which implies an increase in the quantity content of *Spirulina platensis* constituents. Such effect for Cu(II) not observed.

## Introduction

*Spirulina* is a photoautotroph alga and is considered to be a valuable additional food source of some macro- and micronutrients[1]. *Spirulina* contains various components that are beneficial for health, such as proteins, vitamins, essential amino acids, minerals, essential fatty acids. Nowadays, *Spirulina* is cultivated for mass production. There are several studies reporting *Spirulina* therapeutic effects including the hypocholesterolemic[2]. The expression and secretion of inflammatory genes in splenocytes were significantly reduced by *Spirulina* supplementation, demonstrating the anti-inflammatory effects of *Spirulina* in vivo in mouse model[3]. Some environmental stresses, including gamma irradiation and heavy metals, can influence on its. In[4] were investigated the effect of gamma irradiation on the growth, biomass, pigment content and some metabolic activities of *A. platensis*. They reported that gamma irradiation had a stimulatory effect on its growth and cellular constituents. In our works [5,6] effect of Ag(I), Ni(II), Zn(II) ions on irradiated *Spirulina platensis* after three times irradiation and recultivation and also simultaneous effects of Cd(II), Pb(II) ions and γ-irradiation on stability of *Spirulina platensis* intact cells after 7.2 kGy Cs$^{137}$ gamma irradiation were discussed.

In this work, we have studied effect of metal ions Cu(II), Pb(II), Mg(II) on *Spirulina platensis* cells and their constituents after 3-told Cs$^{137}$ gamma irradiation (every case irradiation dose 20 kGy) and recultivation using UV-VIS spectrometry.

## Materials and Methods

*Spirulina platensis* IPPAS B–256 strain was cultivated in a standard Zarrouk [7] alkaline water–salt medium at 34ºC, illumination ~5000lux, at constant mixing in batch cultures [8]. The cultivation of the *Spirulina platensis* intact cells was conducted for 7 days. The growth was measured by optical density by monitoring of changes in absorption at wavelength 560nm using the UV–Visible spectrometer Cintra 10e. The intact cells suspension of *Spirulina platensis* at pH 10.1 in the nutrition medium was used for scanning the absorption spectra from 400 to 800nm. The concentration of *Spirulina platensis* was determined by the instrumental data [9, 10]. *Spirulina platensis* which was irradiated with 7.2 kGy and then recultivated, after one year was re-cultivated again during 7 days. *Spirulina platensis* suspension (100 ml) after 7 days of cultivation have been irradiated with 20kGy gamma radiation using $^{137}$Cs as a gamma radiation source at the Applied Research Center, E. Andronikashvili Institute of Physics (Dose rate -1.1Gy during one minute). Suspension after the irradiation (20 kGy) were filled up to 200ml with Zarrouk medium and they were recultivated. This irradiation and recultivation were repeated 3-told. The optical density was measured every day with 24h intervals. The solutions of metal ions were prepared in deionized water with appropriate amounts of inorganic salt. Reagents $MgCl_2·6H_2O$, $CuCl_2·6H_2O$, $Pb(NO_3)_2$ were of analytical grade. In addition, *Spirulina platensis* image was taken under an electron microscope after irradiation 20kGy and the same cells after 3-told irradiation and re-cultivation.

## Results and Discussions

In fig.1 are presented electron microscope images of intact cells of *Spirulina platensis* (1), cells after irradiation 20kGy (2) and the same cells (3) after 3-told irradiation and re-cultivation. It is clear, that after irradiation the structure of intact cells of *Spirulina platensis* is „broken'' (2) and after three times recultivation the structure was restored (3).

Effect of metals by three times irradiated and recultivated cells of blue-green microalgae *Spirulina platensis* was studied as a function of metal concentration (pH 10.1). Fig.2-4 shows the absorption characteristics of *Spirulina platensis* after three times irradiation (every case irradiation dose 20 kGy) and recultivation and effect of Cu(II), Mg(II), Pb(II) ions on the absorption of *Spirulina platensis*. The peak at 681 nm is due to the absorption of Chl a peak. At 621 nm is due to the absorption of phycocyanin (PC). A peak at 440 nm is due to soret band of Chl a [11]. It is seen from figures, that with increasing metal concentrations absorption intensity decreased for Cu(II), Pb(II) metal ions and increased for Mg(II) ions. For illustration only some absorption spectra are shown in the Fig.2-4 for disregard pictures rebooting and trend readings.

By us were also investigated influence of the same metal ions on the same cellular components of *Spirulina platensis* the same irradiation dose. Effect of metal ions on the absorption intensity maximums for wavelengths 440nm, 621 nm and 681nm are shown in fig.5-7. At 440 nm with increasing Mg(II) concentration the intensity of absorption increases, almost does not change at small concentrations for Pb(II) ions and then for increasing Pb(II) concentrations (4mM) absorption intensity increases. Increasing the concentration for Cu (II) ions causes a decrease in absorption intensity (Fig.5).

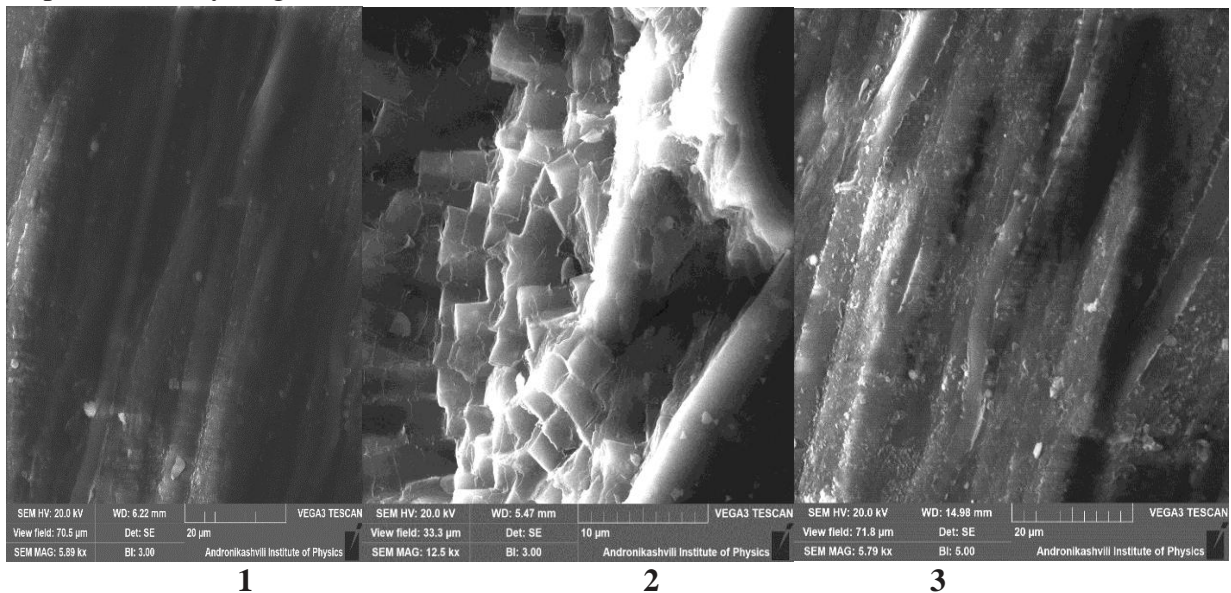

Fig.1. Electron microscope images of intact cells of *Spirulina platensis* (1), cells after γ-irradiation 20kGy (2) and the same cells (3) after 3 -told irradiation and re-cultivation .

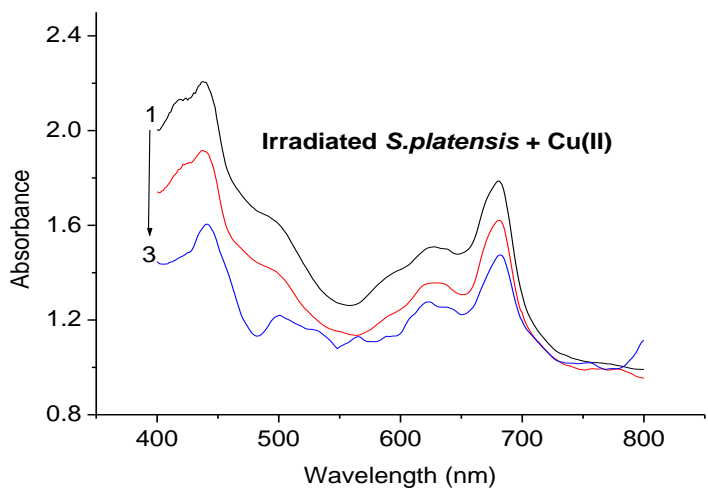

Fig.2. Absorption spectra of *Spirulina platensis* after 3-told γ- irradiation (every case irradiation dose 20 kGy) and re-cultivation-(1), (This cells was irradiated with 7.2 kGy and then recultivated, after one year was re-cultivated, irradiated and re-cultivated again), same irradiated suspension + 1 mMCu(II)-(2), same irradiated suspension + 2mMCu(II)-(3).

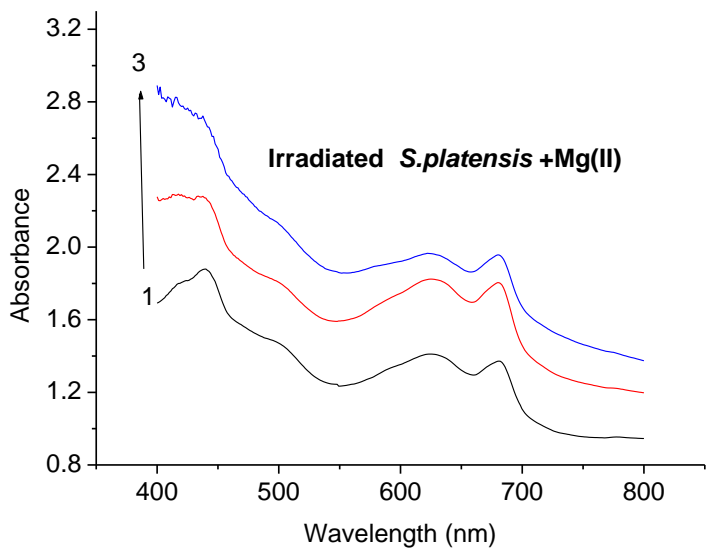

Fig.3. Absorption spectra of the same *Spirulina platensis* after 3-told γ-irradiation and re-cultivation -(1), as for fig.2, same irradiated suspension + 1 mMMg(II) - (2), same irradiated suspension + 2 mM Mg(II)- (3).

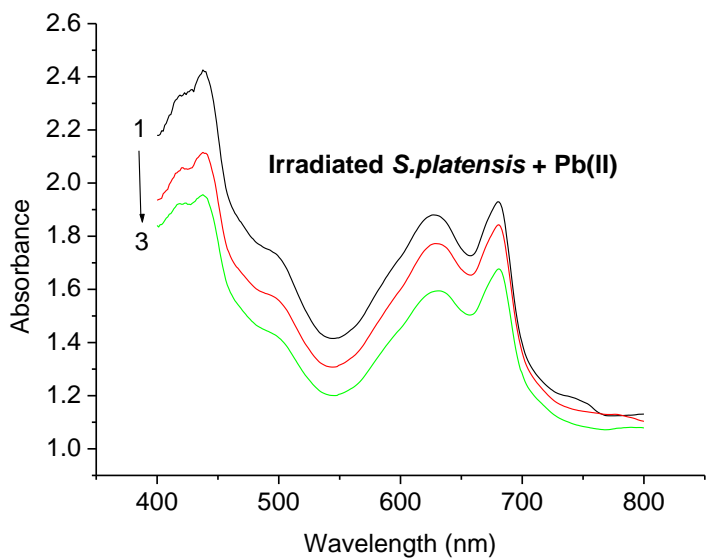

Fig.4. Absorption spectra of the same *Spirulina platensis* after 3-told γ-irradiation and re-cultivation -(1), as for fig.2, same irradiated suspension + 1 mM Pb(II) - (2), same irradiated suspension + 2 mM Pb(II) -(3).

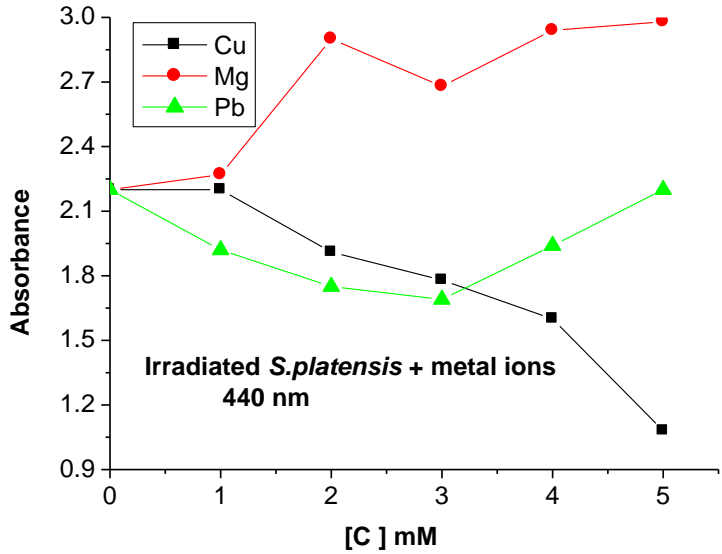

Fig.5. Changes after irradiation in to soret band into the chlorophyll a of *Spirulina platensis* under various metal ions

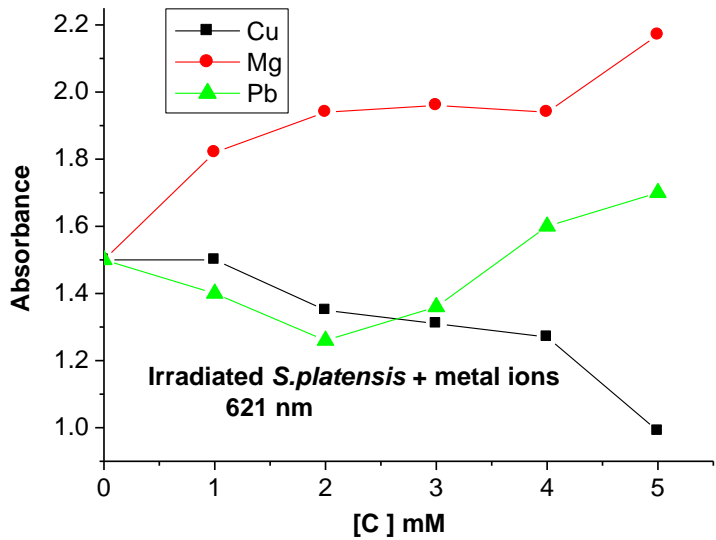

Fig.6. Changes after irradiation in the phycocyanin of *Spirulina platensis* under various metal ions

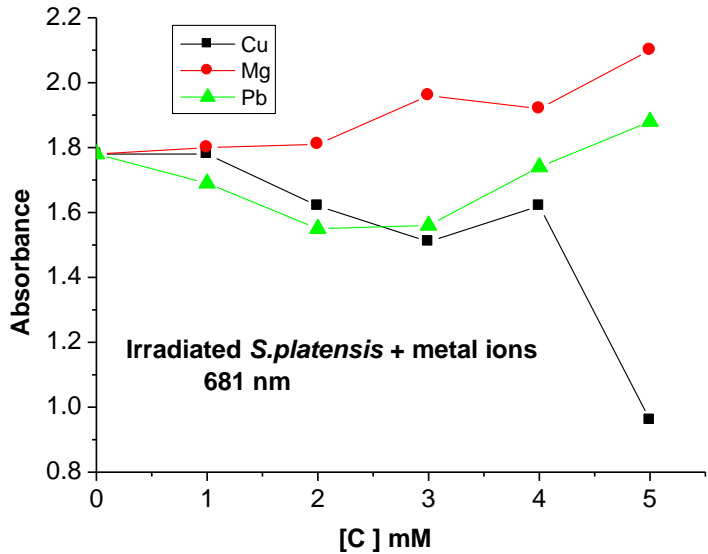

Fig.7. Changes after irradiation in the chlorophyll a of *Spirulina platensis* under various metal ions

As for the change in absorption intensity at 621 nm, which is the peak of absorption of the major protein-phycocyanin of *Spirulina platensis* (Fig.6), as the concentration of Mg(II) ions increases, the absorption intensity very efficiently increases. But unlike Mg(II) ions, an increase in the concentration of copper(II) causes the most noticeable decrease in the intensity of absorption. Absorption intensity almost does not change at small concentrations for Pb(II) ions and then for increasing Pb(II) concentrations (3mM) increases. Similar results were observed in absorption intensity for Mg(II), Cu(II) and Pb(II) ions at 681 nm, which is the absorption peak of the Ch a of *Spirulina platensis*. In particular, that as the concentration of Mg(II) ions increases, the absorption intensity increases very efficiently in the case of irradiation. An increase in the concentration of Cu(II) causes the most noticeable decrease in the intensity of absorption. At 681 nm for Pb(II) ions were observed, changes of the same nature as in the case of 621 nm. Similar results were received in nonirradiation cases (unnpublished results). Based on this by us does not shown absorption spectra for intact cells of *Spirulina platensis*_ + Mg(II), or Cu(II), or Pb(II), ions and effect of these ions on its constituents.

Based on the results obtained, it can be concluded that metal ions Mg(II), Pb(II), Cu(II) differ from each other in terms of interaction efficiency. In particular, influence of Mg(II) ions after several times with discrete doses of radiation and cultivation for *Spirulina platensis* show that as the concentration of Mg(II) ions increases, the absorption intensity very efficiently increases for all constituents. Thus, the combined effects of ionizing radiation and Mg (II) ions for *Spirulina platensis* show synergistic effects for its components as a stimulating agent to increase its content. Whereas in contrast to the Mg(II) ions effect of Cu(II) ions after discrete doses of gamma irradiation and cultivation at increasing the concentration of Cu(II) causes the most noticeable decrease in the intensity of absorption. For Pb(II) ions change of absorption intensity at low concentracions of Pb(II) decreases slowly and then with an increasing concentration of Pb(II) absorption intensity increases for C-phycocyanin, Chl a i.e. the content of all components of *Spirulina platensis* increases in the case of high concentrations of Pb(II).

REFERENCES

1. Ciferri O., *Spirulina*, the edible microorganism. Microbiol Rev. 1983, 47, 551–578.

2. Nagaoka, S., Shimizu, K., Kaneko, H., Shibayama, F., Morikawa, K., Kanamaru, Y., Otsuka, A., Hirahashi, T. & Kato T., A novel protein C-phycocyanin plays a crucial role in the hypocholesterolemic action of *Spirulina platensis* concentrate in rats. J. Nutr. 2005,135, 2425–2430.

3. Tho X. Pham, Yoojin Lee, Minkyung Bae, Siqi Hu, Hyunju Kang, Mi-Bo Kim, Young-Ki Park and Ji-Young Lee, *Spirulina* supplementation in a mouse model of diet-induced liver fibrosis reduced the pro-inflammatory response of splenocytes. British Journal of Nutrition, 2019, 121, 748–755. doi:10.1017/S0007114519000126

4. Effat Fahmy Shabana, Mahmoud AliGabr, Helal RagabMoussa,Enas AliEl-Shaer, Mostafa M.S.Ismaiel, Biochemical composition and antioxidant activities of *Arthrospira* (*Spirulina*) *platensis* in response to gamma irradiation. Food Chemistry, 2017,214, 550-555. doi.org/10.1016/j.foodchem.2016.07.109

5. Gelagutashvili E., Bagdavadze N., Gongadze A., Gogebashvili M., Ivanishvili N.,
Effect of Ag(I), Ni(II), Zn(II) ions on Irradiated *Spirulina platensis*.
arXiv:2102.02007 [physics.bio-ph], 2021.

6. Monaselidze, J., Gelagutashvili,E., Bagdavadze,N., Gorgoshidze, M. and Lomidze,E.,

Simultaneous effects of Cd(II) and Pb(II) ions and γ-irradiation on stability of *Spirulina platensis*. Europ. Chem.Bull. 2019, 8(2), 39-43. doi: 10.17628/ecb.2019.8.38-43

7. Zarrouk C., Contribution to the cyanophyceae study: influence various physical and chimical factors on growth and photosynthesis of *Spirulina maxima* (Setch et Gardner) Geitler extract. Doctorate Thesis, Faculty of Sciences. University of Paris, France. 1966, 146pp.

8. Mosulishvili,L., Belokobilsky, A., Gelagutashvili,E., Rcheulishvili,A., Tsibakhashvili,N.,
 The Study of the mechanism of cadmium accumulation during the cultivation of *Spirulina Platensis*. Proceedings of the Georgian Acad. of Sciences, Biol. series, 1997, 23(1-6), 1997, 105-113.

9. Bennet, A., Bogorad, L.,Complementary chromatic adaptation in a filamentous blue-green alga. *J. Cell Biol.*, **1973**, *58*, 419. doi.org/10.1083/jcb.58.2.419

10. Patel, A., Mishra, S. M., Pawar, R., Ghosh, P., Purification and characterization of C-Phycocyanin from cyanobacterial species of marine and fresh water habitat. *Protein Express. Purif.*, 2005, 40, 248. doi.org/10.1016/j.pep.2004.10.028

11. Lau, R.H., M.M.MacKenzie and W.F. Doolitle., Changes of C-Phycocyanin in *Synechococcus 6301* in Relation to Growth on various Sulfur Compounds. J. Bacteriol. 1979,132,771-778.